\documentclass{article}
\usepackage{dcase2017,amsmath,graphicx,url,times,booktabs, tabularx}
\usepackage{multirow}
\usepackage{color,soul}
\usepackage{flushend}

\title{A report on sound event detection with different binaural features}

\name{Sharath Adavanne, Tuomas Virtanen \thanks{The research leading to these results has received funding from the European Research Council under the European Union’s H2020 Framework Programme through ERC Grant Agreement 637422 EVERYSOUND. The authors also wish to acknowledge CSC-IT Center for Science, Finland, for computational resources.}}
\address{Department of Signal Processing , Tampere University of Technology}

\begin{document}

\ninept
\maketitle

\begin{sloppy}

\begin{abstract}
\end{abstract}
In this paper, we compare the performance of using binaural audio features in place of single channel features for sound event detection. Three different binaural features are studied and evaluated on the publicly available TUT Sound Events 2017 dataset of length 70 minutes. Sound event detection is performed separately with single channel and binaural features using stacked convolutional and recurrent neural network and the evaluation is reported using standard metrics of error rate and F-score. The studied binaural features are seen to consistently perform equal to or better than the single-channel features with respect to error rate metric.

Polyphonic sound event detection, binaural, monochannel, convolutional recurrent neural network

\section{Introduction}
\label{sec:intro}

Sound event detection (SED) is the task of recognizing the sound events and their respective temporal start and end time in a recording. Sound events in real life do not always occur in isolation, but tend to considerably overlap with each other. Recognizing such overlapping sound events is referred as polyphonic SED. Applications of such polyphonic SED are numerous. Recognizing sound events like alarm and glass breaking can be used for surveillance~\cite{surveillance_audio,surveillance}. Environmental sound event detection can be used for monitoring biodiversity studies~\cite{environmentalSED,Marques2012,Furnas2014}. Further, SED can be used for automatically annotating audio datasets, and the sound events recognized can be used as a query for retrieval.

Polyphonic SED using monochannel audio has been studied extensively. Different approaches have been proposed using supervised classifiers like Gaussian mixture model - hidden Markov model~\cite{Mesaros2010_EUSIPCO}, fully-connected networks~\cite{emre2015}, convolutional neural networks (CNN)~\cite{Zhang2015,Phan2016} and recurrent neural networks (RNN)~\cite{giam2016,Adavanne2016_DCASE}. More recently, the state of the art method for polyphonic SED was proposed in~\cite{emre_TASLP2016} and evaluated on multiple private and publicly available datasets. They used log mel-band energies along with a convolutional recurrent neural network (CRNN) architecture as their method.

Recognizing overlapping sound events using monochannel audio is a difficult task. These overlapping sound events can potentially be recognized better with multichannel audio. One of the first methods to use multichannel audio for SED was~\cite{temko2007}. They performed SED on each of the monochannel audio and the combined likelihoods across channels was used for the final prediction. More recently~\cite{Adavanne2017} extended the state of the art CRNN network of~\cite{emre_TASLP2016} for multichannel features and multiple feature classes and showed that using binaural instead of monochannel recordings of the same datasets used in~\cite{emre_TASLP2016} improved the SED performance. Binaural features exploiting the inter-aural intensity and time differences were used in this method. These initial results on binaural audio motivates us to further explore polyphonic SED using binaural audio.

In this paper, we explore and study the performance of three different binaural features - a) log mel band energy, b) log mel band energy extracted in three different resolution windows and c) magnitude and phase component of short-term Fourier transform, all extracted in both the channels of the binaural audio. While a) has been used in~\cite{Adavanne2016_DCASE,Adavanne2017}, b) and c) has not been used in polyphonic SED task previously. We compare the performance of SED amongst the binaural features and also compare with the single channel log mel-band energy feature. We separately train the multichannel network method~\cite{Adavanne2017} with features extracted from the publicly available TUT Sound Events 2017 dataset and present the results.

The feature extraction and neural network used is described in section~\ref{sec:method}. The dataset creation, evaluation metrics and procedure are explained in section~\ref{sec:eval}. Finally, the results and discussion are presented in section~\ref{sec:results}.

\begin{figure}[!ht]
  \centering
  \centerline{\includegraphics[width=\columnwidth]{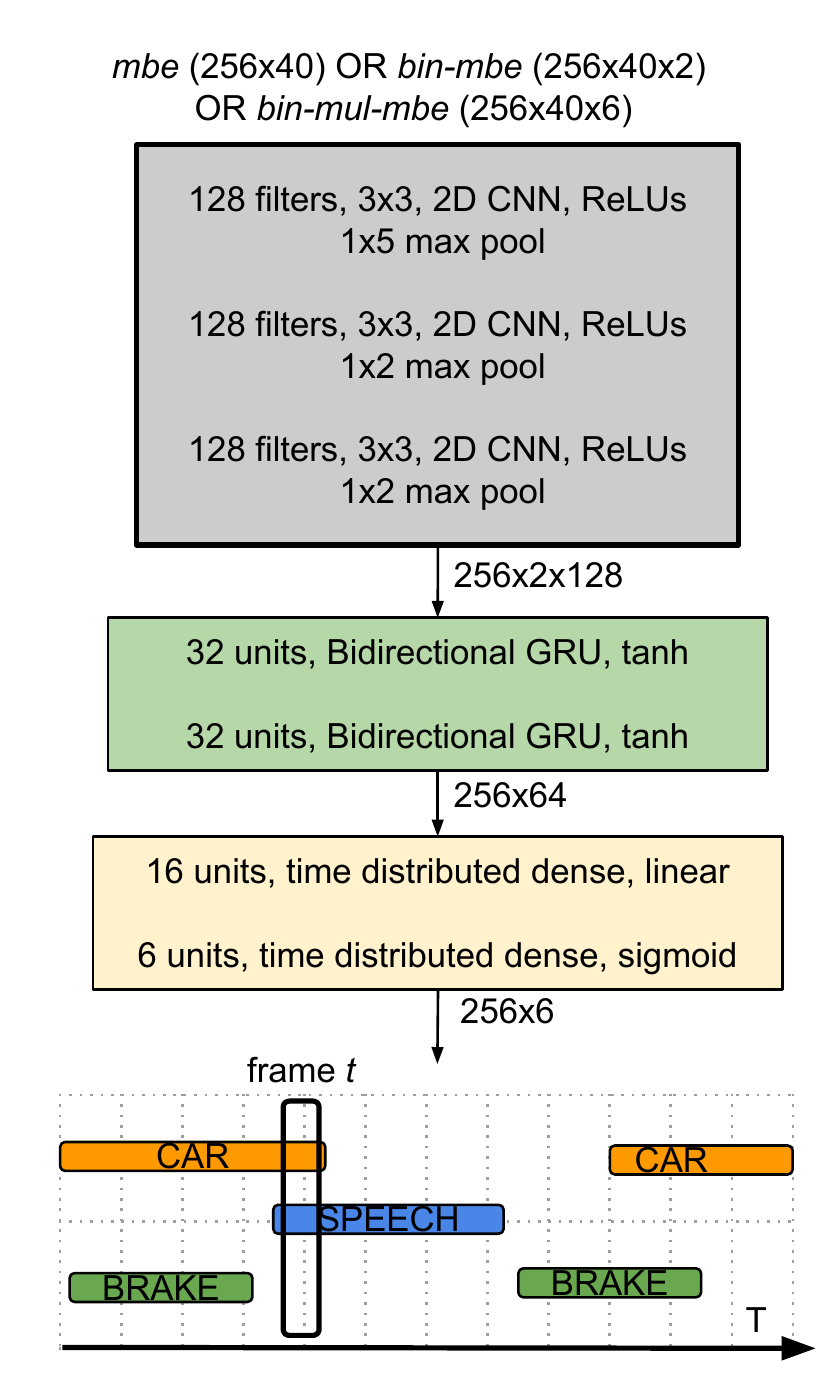}}
  \caption{Stacked convolutional and recurrent neural network for binaural polyphonic sound event detection.}
  \label{fig:crnn}
\end{figure}

\section{Method}
\label{sec:method}
The input to the method is an audio signal. Features are extracted in consecutive time windows from each channel of the audio. These audio features are fed to a multichannel convolutional and recurrent neural network architecture, which maps the audio features to the sound event labels in the dataset. The output of the neural network is in the range of [0, 1] for each of the sound event label, where one refers to the sound event being active, and zero for absence. The detailed description of feature extraction and the neural network is presented below.

\subsection{Feature extraction}
\label{ssec:feat}
In this paper, we study the performance of three binaural audio features and compare it with single channel audio feature. All features are extracted in hop length of 20 ms to keep the number of frames same.

\subsubsection{Single channel feature}
Log mel-band energy ($mbe$) has been used extensively for the SED task~\cite{giam2016,emre_TASLP2016,Adavanne2016_DCASE,Adavanne2017} we continue to use the feature in this paper. $mbe$ is extracted in Hamming window of length 40 ms. We use 40 mel-bands in the frequency range of 0-22500 Hz. For a given audio input of $F$ frames, this feature extraction block results in a $F\times40$ output.

\subsubsection{Binaural features}
The first binaural feature we study is the binaural log mel-band energy ($bin$-$mbe$) from the works of~\cite{Adavanne2016_DCASE}, where it was shown to perform better than the $mbe$. We extract $bin$-$mbe$ in a similar fashion as $mbe$ on each of the binaural channels resulting in a $F\times80$ (40*2 = 80) output.

Su et al. in ~\cite{Su2017} reported that $mbe$ extracted in multi-resolution windows give considerable improvement for SED over using just the single resolution $mbe$. Motivated by this we extend it to binaural scenario, and extract it in both the channels of audio ($bin$-$mul$-$mbe$). Specifically, we use three different window sizes 1024, 4096, and 16384 as in the paper~\cite{Su2017} and extract $mbe$ feature in each of the windows and each of the binaural channels. This feature extraction block results in a $F\times240$ (40*3*2 = 240) output. 

Recently, it was shown that the neural networks can estimate the direction of arrival from just the phase components of the multi-channel short-term Fourier transform (STFT) coefficients~\cite{Chakrabarty2017}. Motivated by this, we extend it to binaural channels by extracting STFT in each of the binaural channels and propose to also use the magnitude component along with the phase component ($bin$-$fft$). We extract STFT in windows of 40 ms using 2048 points, post which we calculate the magnitude and the phase component resulting in a $F\times4096$ (1024*2*2 = 4096) output. 

\subsection{Neural network}
\label{ssec:dnn}

The input to the neural network in the case of single channel audio is $T$ consecutive time frames of $mbe$, with a dimension of $T\times40$ as shown in Figure~\ref{fig:crnn}. In the case of binaural audio features, we stack each of the channel features separately. Specifically, for $bin$-$mbe$ we stack the $T\times80$ output of feature extraction block to a volume of dimension $T\times40\times2$. In case of $bin$-$mul$-$mbe$ and $bin$-$fft$ along with the channels, we also stack the multi-resolution windows, phase and magnitude components separately, resulting in volumes of dimension $T\times40\times6$ and $T\times1024\times4$ respectively. Based on the task of mono or binaural SED, the network is fed with the respective feature sequence. In this paper, we use a sequence length of T = 256 for all the features.

We use convolutional neural network (CNN) as our initial layers to learn local shift-invariant patterns from audio feature. The receptive filters of these CNNs are of the size $3\times3$ size. The output activation from the CNN layers are padded with zeros to keep the dimension of the output the same as input. Batch normalization~\cite{batchNorm} and max-pooling is performed after every layer of CNN to reduce the final dimension to $T\times2\times N$, where N is the number of filters in the final layer of CNN. We perform max-pooling in the frequency axis only, this is done to preserve the time resolution of the input. The CNN layer activation is further fed to layers of bi-directional gated recurrent units (GRU), to learn long term temporal activity patterns. This is followed by layers of time-distributed fully-connected (dense) layers. The time resolution remains the same as input feature in both GRU and dense layers. The final prediction layer has an output dimension of $T\times C$, where $C$ is the number of classes in the dataset. The prediction layer has sigmoid activation in order to be able to produce multi-label output.

The training is performed for 500 epochs using binary cross-entropy loss function and Adam~\cite{adamKeras} optimizer with a learning rate of 0.0001. Dropout~\cite{Dropout} is used as a regularizer after every layer of neural network to make it robust to unseen data. Early stopping is used to stop over-fitting the network to training data. Training is stopped if the error rate (see Section~\ref{ssec:metric}) on the test split does not improve for 100 epochs. The neural network implementation was done using Keras~\cite{chollet2015keras} framework with Theano~\cite{theano} as backend.

\begin{table}[!ht]
\centering
\begin{tabular}{l|l}
Sound events & Length \\ \hline
brakes squeaking & 67.6 \\
car & 2541.5 \\
children & 346.1 \\
large vehicle & 727.0 \\
people speaking & 630.6 \\
people walking & 1079.2 \\ \hline
Total & 5391.9
\end{tabular}
\caption{Distribution of sound event in the dataset. The length is given in seconds.}\vspace{-10pt}
\label{T:dataset}
\end{table}

\section{Evaluation} 
\label{sec:eval}

\subsection{Dataset}
\label{ssec:data}
We study the performance of the binaural features on the development set of TUT sound events 2017 dataset organized as part of Detection and Classification of Acoustic Scenes and Events~\cite{dcase2017}.

This dataset consists of about 70 minutes of audio data collected in street scenario with annotations of six sound event classes. Table~\ref{T:dataset} presents the sound event classes and their distribution in the dataset. There are 24 recordings in total, each of about 3-5 minutes, recorded using  Soundman OKM II Klassik/studio A3 electret in-ear microphone and a Roland Edirol R-09 wave recorder. The recordings are sampled at 44.1 kHz and 24 bit resolution. The single channel audio for single channel feature study is created by taking the average of the binaural channel audio. The dataset provides four cross validation splits for the above data, with train, validation and test splits.

\subsection{Metric}
\label{ssec:metric}
The SED method is evaluated using the polyphonic SED metrics proposed in~\cite{metrics}. Particularly we use segment wise error rate (ER) and F-score calculated in segments of one second length. According to which the F-score is calculated as, 
\begin{equation}
F = \frac{2 \cdot \sum_{k=1}^{K} TP(k)}{2 \cdot \sum_{k=1}^{K}TP(k)+ \sum_{k=1}^{K}FP(k)+ \sum_{k=1}^{K}FN(k)},
\end{equation}
where for each one second segment $k$,
$TP(k)$ is the true positives, the number of sound event labels active in both predictions and groundtruth. $FP(k)$ is the false positives, the number of sound event labels active in predictions but inactive in groundtruth. $FN(k)$ is the false negatives, the number of sound event labels active in the ground truth but inactive in the predictions.

The error rate is measured as,
\begin{align}
ER = \frac{\sum_{k=1}^{K} S(k) + \sum_{k=1}^{K} D(k) + \sum_{k=1}^{K} I(k)}{\sum_{k=1}^{K} N(k)},
\end{align}
where, $N(k)$ is the total number of active sound events in the ground truth of segment $k$. The  substitutions ($S(k)$), deletions ($D(k)$) and insertions($I(k)$) are measured using the following equations for each of the $K$ one second segments.

\begin{align}
S(k) = \min(FN(k), FP(k)) \\
D(k) = \max(0, FN(k)-FP(k)) \\
I(k) = \max(0, FP(k)-FN(k)) 
\end{align}

For an ideal SED method, ER is zero and F-score is 100.

\subsection{Baseline}
\label{ssec:baseline}
The baseline method for the dataset used is provided in~\cite{dcase2017}. This method uses single channel $mbe$ as the audio feature. The network consists of two fully-connected layers with 50 units in each followed by a dropout layer with 0.2 dropout rate. The prediction layer has number of sigmoid units equal to the number of classes in the dataset. The method uses a context of 5 frames resulting in a feature length of 200 (40*5). The network is trained with cross-entropy loss and Adam optimizer for 200 epochs. The evaluation metric scores for this method is reported in Table~\ref{T:results}.

\subsection{Evaluation procedure}
\label{ssec:eval_proc}
A random hyper-parameter search~\cite{Bergstra2012} is performed by varying the number of layers and units of CNN, GRU and dense layers, and the dropout in the set of \{0.05, 0.25, 0.5, 0.75\} for each of the feature. The hyper-parameter tuning was done to achieve the best ER on the test split. The best configuration found was the same for all the mel based features ($mbe$, $bin$-$mbe$, $bin$-$mul$-$mbe$), while the $bin$-$fft$ stereo was seen to give good results with the same network but larger max-pooling ($1\times8$ after each layer of CNN).  The network for mel based features and its configuration is as shown in Figure~\ref{fig:crnn}. The dropout rate for the above configuration was 0.5 for $mbe$ and $bin$-$mbe$, 0.25 for $bin$-$mul$-$mbe$ and 0.05 for $bin$-$fft$

We perform SED on the above dataset individually with all the single channel and binaural features and report the average ER and F-scores of five separate runs of four cross-validation provided in the dataset.

\begin{table}[!htb]
\centering
\begin{tabular}{l|cc|cc}
 & \multicolumn{2}{c|}{Development} & \multicolumn{2}{c}{Challenge} \\ \cline{2-5}
Audio features & ER & F& ER & F \\ \hline
baseline ~\cite{dcase2017} & 0.69 & 56.7 & 0.94& 42.8\\ 
$mbe$            & 0.55& 69.3& \bf{0.79}& 41.7\\ \hline
$bin$-$mbe$        & 0.52& 69.1& 0.80& \bf{42.9}\\
$bin$-$mul$-$mbe$    & \bf{0.50}& \bf{70.3}& 0.85& 41.4\\
$bin$-$fft$           & 0.55& 66.9& 0.87& 36.2      
\end{tabular}
\caption{Best evaluation metric scores achieved with different audio features on the development dataset and the evaluation dataset of DCASE 2017 challenge~\cite{dcase2017}.}
\label{T:results}
\end{table}

\section{Results and discussion} 
\label{sec:results}
The evaluation results for SED using single channel and binaural features are  presented in Table~\ref{T:results}. We see that the stacked convolutional and recurrent neural network with the single channel audio feature ($mbe$) outperforms the baseline method~\cite{dcase2017}. 

Binaural features in general have similar performance as single channel features on the evaluated dataset. In particular the ER of binaural features is seen to be equal or better than the single channel feature. The noteworthy performance is of $bin$-$mul$-$mbe$ which is seen to improve the ER considerably over $mbe$. 

The validation and training loss of the network with $bin$-$fft$ was considerably higher than the other features. Suggesting that the size of the data was possibly less for this feature to find the best weights.

\subsection{DCASE 2017 challenge results}
The stacked convolutional and recurrent neural network trained respectively with the three binaural and the single channel feature was submitted in the real life sound event detection task of DCASE 2017 challenge~\cite{dcase2017}. The results obtained on the evaluation data of the challenge is presented in Table~\ref{T:results}. All the submitted systems fared well on the evaluation data and resulted in challenging scores. In particular, the network trained on $mbe$ fared as the best method in the challenge followed by $mbe$-$bin$ in close second among the 34 submitted methods from 14 different teams.

\section{Conclusion} 
\label{sec:conclusion}
In this paper, we proposed to study the performance of using different binaural audio features for polyphonic sound event detection. In this regard three binaural features were studied and compared with a baseline single channel audio feature. We performed SED separately for each of the feature using a stacked convolutional and recurrent neural network. The evaluation was carried out on the publicly available TUT Sound Events 2017 dataset. It was observed that using binaural features gave similar or better error rate than single channel features. In particular log mel-band energy feature extracted in different resolution windows was seen to produce the best results for the given dataset.

\bibliographystyle{IEEEtran}
\bibliography{refs}
\end{sloppy}
\end{document}